\documentclass[12pt,preprint]{aastex}

\shorttitle{Sub-mm detections of Spitzer populations}
\shortauthors{Serjeant et al.}

\begin{document}


\title{Sub-millimeter detections of 
{\it Spitzer Space Telescope} galaxy populations}



\author{
S.\ Serjeant,\altaffilmark{1}
A.\,M.\,J.\ Mortier,\altaffilmark{1}
R.\,J.\ Ivison,\altaffilmark{2} 
E.\ Egami,\altaffilmark{3}
G.\,H.\ Rieke,\altaffilmark{3}
S.\,P.\ Willner,\altaffilmark{4}
D.\ Rigopoulou,\altaffilmark{6}
A.\ Alonso-Herrero,\altaffilmark{3}
P.\ Barmby,\altaffilmark{4}
L.\ Bei,\altaffilmark{3}
H.\ Dole,\altaffilmark{3,8}
C.\,W.\ Engelbracht,\altaffilmark{3}
G.\,G.\ Fazio,\altaffilmark{4}
E.\ Le Floc'h,\altaffilmark{3}
K.\,D.\ Gordon,\altaffilmark{3}
T.\ R.\ Greve,\altaffilmark{5}
D.\,C.\ Hines,\altaffilmark{3,10}
J.\,-S.\ Huang,\altaffilmark{4}
K.\,A.\ Misselt,\altaffilmark{3}
S.\ Miyazaki,\altaffilmark{9}
J.\,E.\ Morrison,\altaffilmark{3}
C.\ Papovich,\altaffilmark{3}
P.\,G.\ P\'{e}rez-Gonz\'{a}lez,\altaffilmark{3}
M.\,J.\ Rieke,\altaffilmark{3}
J.\ Rigby,\altaffilmark{3}
G.\ Wilson\altaffilmark{7}
}

\altaffiltext{1}{Centre for Astrophysics \& Planetary Science, School
of Physical Sciences, University of Kent, Canterbury, Kent, CT2 7NR, UK}
\altaffiltext{2}{Astronomy Technology Centre, Royal Observatory,
Blackford Hill, Edinburgh, EH9 3HJ, UK} 
\altaffiltext{3}{Steward Observatory, University of Arizona, 933 N
Cherry Avenue, Tuscon, AZ, 85721}
\altaffiltext{4}{Harvard-Smithsonian Center for Astrophysics, 60
Garden Street, Cambridge, MA 02138}
\altaffiltext{5}{Institute for Astronomy, University of Edinburgh,
Blackford Hill, Edinburgh, EH9 3HJ, UK}
\altaffiltext{6}{Astrophysics, Denys Wilson Building, Keble Road,
Oxford, OX1 3RH, UK}
\altaffiltext{7}{Spitzer Science Center, California Institute of
Technology, Mail Code 220-6, 1200 East California Boulevard, Pasadena,
CA 91125}
\altaffiltext{8}{IAS, bat 121, Universite Paris-Sud,
F-91405 Orsay cedex, France}
\altaffiltext{9}{Subaru Telescope, National Astronomical Observatory
of Japan, 650 North A'ohoku Place, Hilo, HI 96720}
\altaffiltext{10}{Space Science Institute, 4750 Walnut Street, Suite
205, Boulder, Colorado 80301}
\setcounter{footnote}{11}

%
%
%



\begin{abstract}
We present 
sub-millimeter statistical detections of
galaxies discovered in the $5'\times5'$ Spitzer Early Release
Observations (to $\sim4-15\mu$Jy $5\sigma$ at $3.6-8\mu$m,
$170\mu$Jy at $24\mu$m)
through a stacking analysis of 
our reanalysed SCUBA 8mJy survey maps, and a Spitzer 
identification of a new sub-millimeter
point source in the 8mJy survey region. 
For sources detected at $5.8\mu$m or $8\mu$m ($154$ and $111$ sources 
respectively),  
we detect positive skews in the sub-millimeter flux distributions 
at $99.2-99.8\%$ confidence using Kolmogorov-Smirnov tests, at both
$850\mu$m and $450\mu$m. We also marginally detect the {\it Spitzer}
$24\mu$m galaxies at $850\mu$m at $97\%$ confidence, and place
limits on the mean sub-millimeter fluxes of the $3.6\mu$m and
$4.5\mu$m sources. 
Integrating the sub-millimeter fluxes of the {\it Spitzer}
populations, we find the $5.8\mu$m galaxies 
contribute 
$0.12\pm0.05$ nW m$^{-2}$ sr$^{-1}$ to the
$850\mu$m background, and 
$2.4\pm0.7$ nW m$^{-2}$ sr$^{-1}$ to the
$450\mu$m background; similar contributions 
are made by the $8\mu$m-selected sample. 
We infer that the populations dominating the $5.8\mu$m and $8\mu$m
extragalactic background light also contribute around a quarter of 
the $850\mu$m background and the majority of the $450\mu$m background. 
\end{abstract}




\keywords{cosmology: observations --- galaxies: evolution ---
galaxies: formation} 


\section{Introduction}

The early pioneering sub-millimeter surveys in lensing clusters
(Smail, Ivison \& Blain 1997)
and in blank fields (Hughes et al. 1998, Barger et al. 1998)
demonstrated the feasibility 
of deep extragalactic surveys
exploiting the favorable K-corrections in the sub-millimeter. 
Deep $850\mu$m imaging has now resolved around half 
of the $850\mu$m extragalactic background (Hughes et
al. 1998, Blain et al. 1999, Cowie, Barger \& Kneib 2002). 
These galaxies are 
often called SCUBA galaxies after the instrument with which they were
first detected (Holland et al. 1999). 
The $14''$ $850\mu$m SCUBA beam makes identifications at other
wavelengths difficult; nevertheless, $\sim50\%$ of $850\mu$m sources are
identifiable in $\mu$Jy $1.4$GHz imaging (Ivison et al. 2002). These radio
identifications have led to 
optical identifications, morphologies and ultimately spectroscopic
redshifts in multiwavelength follow-up campaigns (e.g. Chapman et
al. 2003). Furthermore, 
the non-detection of SCUBA galaxies in hard X-ray imaging
(e.g. Alexander et al. 2003) 
suggests that the bulk of the population has far-infrared luminosities
dominated by star formation. The morphologies, redshifts, 
clustering and molecular gas contents are so far consistent with at
least some of the SCUBA population being the progenitors of giant
elliptical galaxies (e.g. Dunlop 2002), though other
alternatives are still viable (Efstathiou \& Rowan-Robinson 2003) and
the SCUBA population is heterogeneous (e.g. Ivison et
al. 1998, 2000). 
Finally, the K-correction effects in the
sub-millimeter make the sub-millimeter extragalactic background
sensitive to contributions from the far-infrared luminous energy
densities at all redshifts $z\stackrel{<}{_\sim}10$. The
populations which contribute to the $850\mu$m extragalactic background
are necessarily also significant contributors to the 
cosmic history of dust-shrouded star formation. 


Following the {\it IRAS} mid-infrared surveys of the local Universe
(e.g. Rush, Malkan \& Spinoglio 1993), the mid-infrared was first made
accessible 
to deep extragalactic surveys by the Infrared Space Observatory ({\it ISO},
Kessler et al. 1996) which conducted a suite of surveys with a variety
of depths and areal coverages (e.g. Genzel \& Cesarsky 2000 and
refs. therein, Rowan-Robinson et al. 2004). The
rapid upturn in the $15\mu$m extragalactic source counts clearly
demonstrated the existence of a strongly evolving population of
obscured starbursts and active galaxies (e.g. Serjeant et al. 2000, 
Franceschini et
al. 2001, Gruppioni et al. 2002). It has also been 
argued that the populations dominating the $15\mu$m extragalactic
background light, which are resolved by {\it ISO}, are also largely the same
populations which dominate the unresolved $140\mu$m background
(Elbaz et al. 2002). If correct, this is a significant
breakthrough in determining the populations which dominated the
far-infrared luminous energy density throughout the relatively 
recent history of the Universe (e.g. $z<1$). 

Finding the population that supplies the luminous energy density 
at $z>1$ requires understanding the sub-millimeter background light. 
However, 
it has been difficult to find sub-millimeter source counterparts in
the mid-infrared. 
Very few sub-millimeter-selected sources have
been detected by {\it ISO} in the mid-infrared
(e.g. Eales et al. 2000, Webb et al. 2003b, Sato et al. 2002).
The reverse procedure of looking for mid-infrared sources in the
sub-millimeter via stacking analyses have not fared much better. 
Serjeant et al. (2003a) found no excess $850\mu$m flux at the
locations of $15\mu$m sources in the HDF North. 
Lyman break galaxies, in contrast, are detectable statistically
(e.g. Peacock et al. 2000, Webb et al. 2003a). If SCUBA galaxies 
are extreme star-forming galaxies in the most massive high-redshift
halos, 
then their anomalously faint K-band identifications imply heavy
obscuration in the observed-frame near-infrared
(Serjeant et al. 2003b), suggesting that SCUBA galaxies
may be detectable in $\mu$Jy-level mid-infrared imaging. 

The {\it Spitzer Space Telescope} (hereafter {\it Spitzer}, Werner et
al. 2004) is an 
enormous advance over {\it ISO} in terms 
of mapping speed, sensitivity, and wavelength coverage. 
It may now be possible to resolve the bulk of the 
extragalactic background light at
$3.6-8\mu$m in exposures of the order $15$ minutes, an equivalent 
depth achieved at $6.7\mu$m in $23$ hours with {\it ISO} by Sato et
al. 2003. In this paper we present statistical sub-millimeter
detections of galaxies selected in a new {\it Spitzer} survey. 
The {\it Spitzer} identifications of previously-published
sub-millimeter 
sources are discussed by Egami et
al. (2004). 
Identifications of MAMBO sources are discussed by Ivison
et al. (2004). 


%
%
%
\section{Observations}
The {\it Spitzer} Early Release Observations survey is 
one of the first extragalactic surveys conducted by {\it
Spitzer}. Besides the pragmatic goal of 
characterising the survey capabilities of the facility, the
survey has the science goals of making the first constraints on the
populations which dominate the extragalactic backgrounds in the
shorter wavelength {\it Spitzer} bands, and the links between these
galaxies and other known populations. 
Accordingly, the survey field was selected to lie in
the Lockman Hole, an area with abundant multi-wavelength survey coverage,
and in particular with $7$ galaxies from the 
$850\mu$m 8mJy survey $>3.5\sigma$ catalog
(Scott et al. 2002, Fox et al. 2002, Ivison et al. 2002; see below). 

The {\it Spitzer} imaging is described by Egami et al. (2004) and
Huang et al. (2004). In summary, IRAC (Fazio et al. 2004) imaged a 
$\sim5'\times5'$ field  
for $750$s per sky pixel at all four bands, resulting in median
$1\sigma$ depths of $0.77\mu$Jy, $0.78\mu$Jy, $2.75\mu$Jy and
$1.67\mu$Jy at $3.6\mu$m, $4.5\mu$m, $5.8\mu$m and $8\mu$m
respectively, and sources were extracted to $>4.7\sigma$. MIPS
(Rieke et al. 2004) 
observed the field in photometry mode 
at $24\mu$m for $250$s per sky pixel, resulting in a typical $1\sigma$
depth of $30\mu$Jy. Source confusion and blending make the MIPS
completeness and reliability more problematic than for SCUBA or IRAC.
MIPS source catalogs were
extracted to $>170\mu$Jy ($80\%$ complete, $>90\%$ reliable) and 
to $>120\mu$Jy ($50\%$ complete, $50\%$ reliable). We conservatively
assume the catalogs are all extragalactic, as expected at these flux
densities; the deletion of any (hypothetical) 
contaminant population of galactic
stars with negligible sub-millimeter fluxes
would improve the confidence levels of our statistical detections. 

The SCUBA 8mJy survey data (which covers all our {\it Spitzer} field) 
is described in detail by Scott et
al. (2002). We have reanalysed this data, modelling the intra-night
SCUBA gain variations, improving the extinction corrections using
improved fits to 
the $225$GHz skydips monitored at
the Caltech Sub-millimeter Observatory, and removing the cross-talking
$450\mu$m bolometers A7 and A16 (see Mortier et
al. 2004). Additional sources are detectable in the 
zero-sum chopped/nodded 
sub-millimeter maps. After point source detection using 
noise-weighted PSF convolution 
(Serjeant et al. 2003a)
including the effect of negative
chop/nod positions, 
the median $1\sigma$ depth (total flux) in the {\it Spitzer} Early
Release Observations field is $1.6$mJy at $850\mu$m, and $14.4$mJy at
$450\mu$m. 

\section{Results}
\label{sec:ids}
\begin{figure*}[tbh]
\centerline{
\includegraphics*[scale=0.65,angle=0]{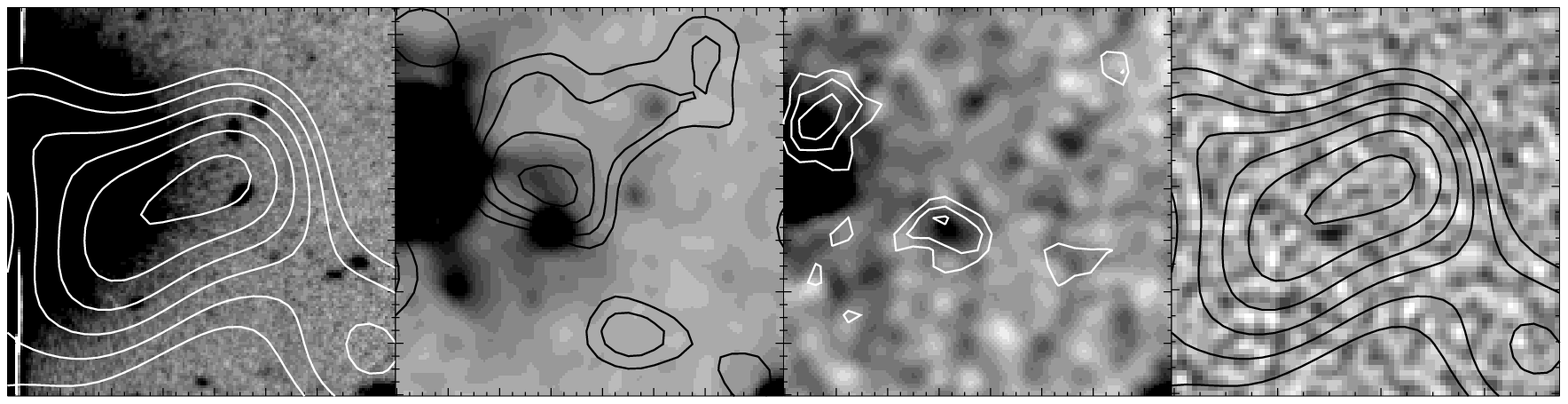}
}
\caption{\small\label{fig:ids} {\it Spitzer} identification of the
sub-millimeter source in section \ref{sec:ids}. 
All postage stamps are $30''$ on a side, with North up and
East to the left. From left to right, the greyscale images show the
Subaru R-band image; $4.5\mu$m; $5.8\mu$m and
VLA $1.4$GHz A-array. 
The contours, from left to right, are $850\mu$m, 
$450\mu$m, $24\mu$m and $850\mu$m. In all cases the contours 
are from $1\sigma$ in steps of $0.5\sigma$. 
}
\end{figure*}

Our reanalysis of the SCUBA 8mJy survey uncovered new candidate
sub-millimeter sources 
(Mortier et
al. 2004). Figure \ref{fig:ids} shows the {\it Spitzer}
identifications of the most significant of these, at $\alpha=10^{\rm
h}$ $51^{\rm m}$ $53.94^{\rm s}$, $\delta=+57^\circ$ $25'$
$05.5''$ (J2000),
detected with 
a $3.7\sigma$ $850\mu$m flux
of $4.3$mJy. At $450\mu$m the source is not detected, though
the $450\mu$m flux at its position is $17.9$mJy ($1.6\sigma$); there
is also a hint of positive flux at $1.8\sigma$ significance at
$1250\mu$m (Greve et al. 2004) at the position of this source. 
There are hints that the source is extended at $850\mu$m, suggesting a
blend of more than one source, and indeed there are two candidate
identifications in the {\it Spitzer} imaging. 
The probabilities of a
random association ($P=1-\exp(-n(>s)\pi r^2)$ where $n(>s)$ is the
surface density of catalogued objects brighter than the 
identification at distance $r$) are $0.033$ and $0.155$ for the
brighter and fainter $4.5\mu$m flux respectively.
Interestingly, the
low-significance contours of $450\mu$m emission are coincident with
the brighter of the two candidates. 
The {\it Spitzer} fluxes of this source
are $16.6$, $26.7$, $14.9$, $18.9\mu$Jy and $216\mu$Jy at $3.6\mu$m, 
$4.5\mu$m, $5.8\mu$m, $8\mu$m and $24\mu$m respectively. The fluxes of
the second 
candidate identification are $6.2\mu$Jy and $4.6\mu$Jy at $3.6\mu$m,
and $4.5\mu$m, but the source is not detected at longer wavelength
{\it Spitzer} bands. While deblending the fluxes from the two
identifications is beyond the scope of this paper, it is interesting
to note that either identification would have a sub-millimeter:{\it
Spitzer} flux ratio which is redder than 
in the population as a whole, derived below. 

The $>3.5\sigma$ sub-millimeter sources are already known to have {\it
Spitzer} identifications (Egami et al. 2004),
so 
we
construct a mask to remove the sources at these positions from
the stacking analysis. 
At $850\mu$m 
we masked a $9.9''$ radius ($\sqrt{2}\times$ the SCUBA FWHM, i.e. the
FWHM of point sources in our PSF-convolved maps) 
around the  $>3.5\sigma$ $850\mu$m point
sources from Scott 
et al. (2002), as well as the new  point sources from our
reanalysis (Mortier et al. 2004). 
At 
$450\mu$m 
the masking radius was $5.2''$,
which we also applied to 
new $450\mu$m sources 
from our reanalysis.
We also masked regions with high noise
levels, arbitrarily selected as 
$>5$mJy at $850\mu$m, or $>100$mJy at $450\mu$m, and
also masked all regions in the sub-millimeter maps 
without {\it Spitzer} data.
The unmasked area at $850\mu$m ($450\mu$m) is $22.1$ 
($22.7$) arcmin$^2$ for the IRAC catalogues. For MIPS, the
corresponding area is $27.6$ ($28.5$) arcmin$^2$. 

\begin{figure*}[tbh]
\centerline{
\includegraphics*[scale=0.35]{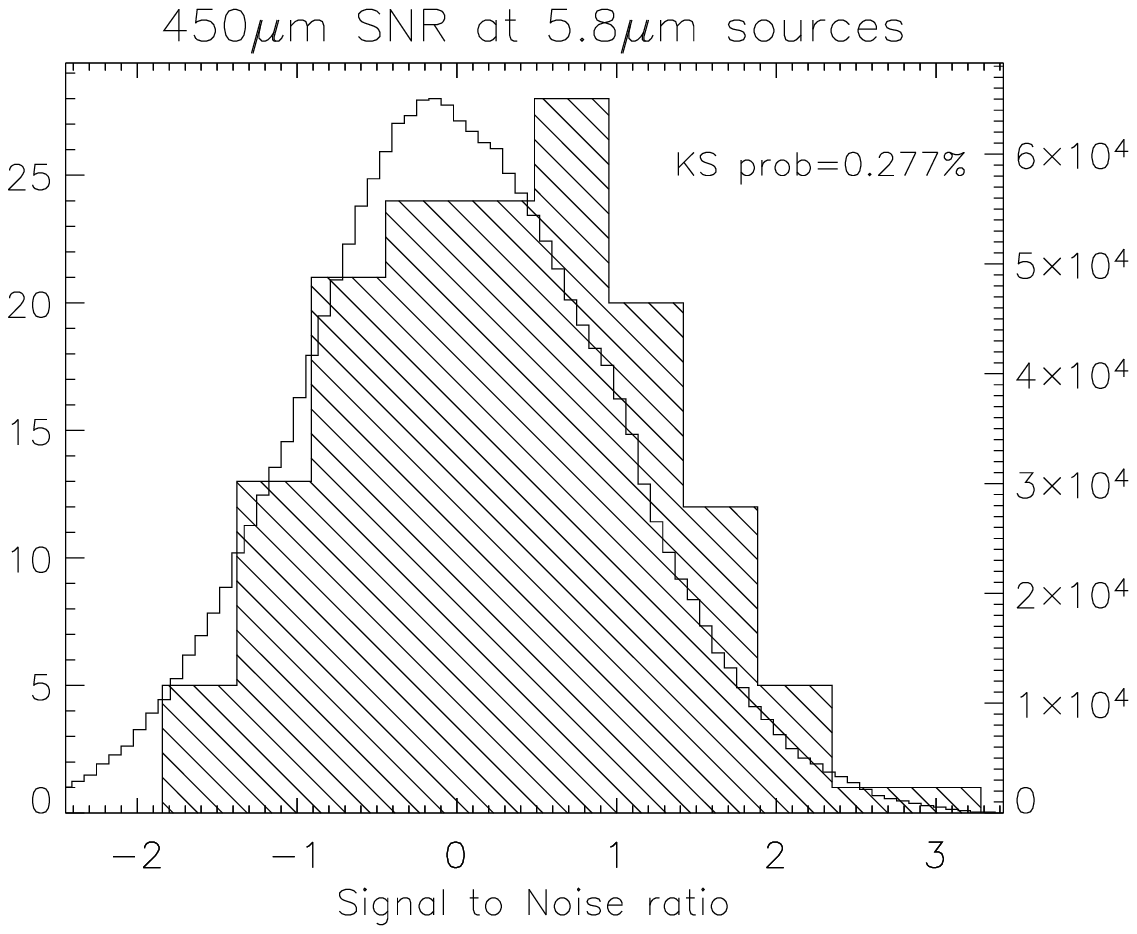}
\includegraphics*[scale=0.35]{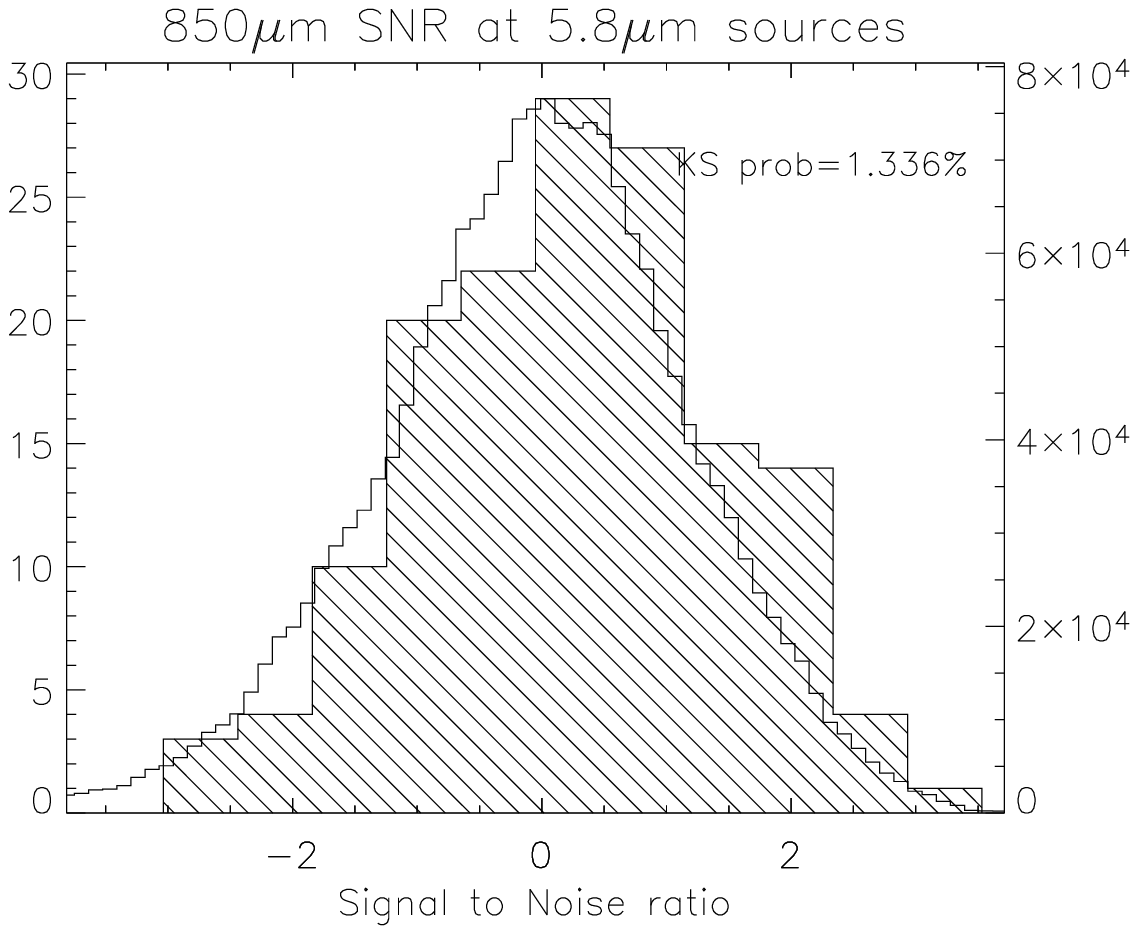}
\includegraphics*[scale=0.35]{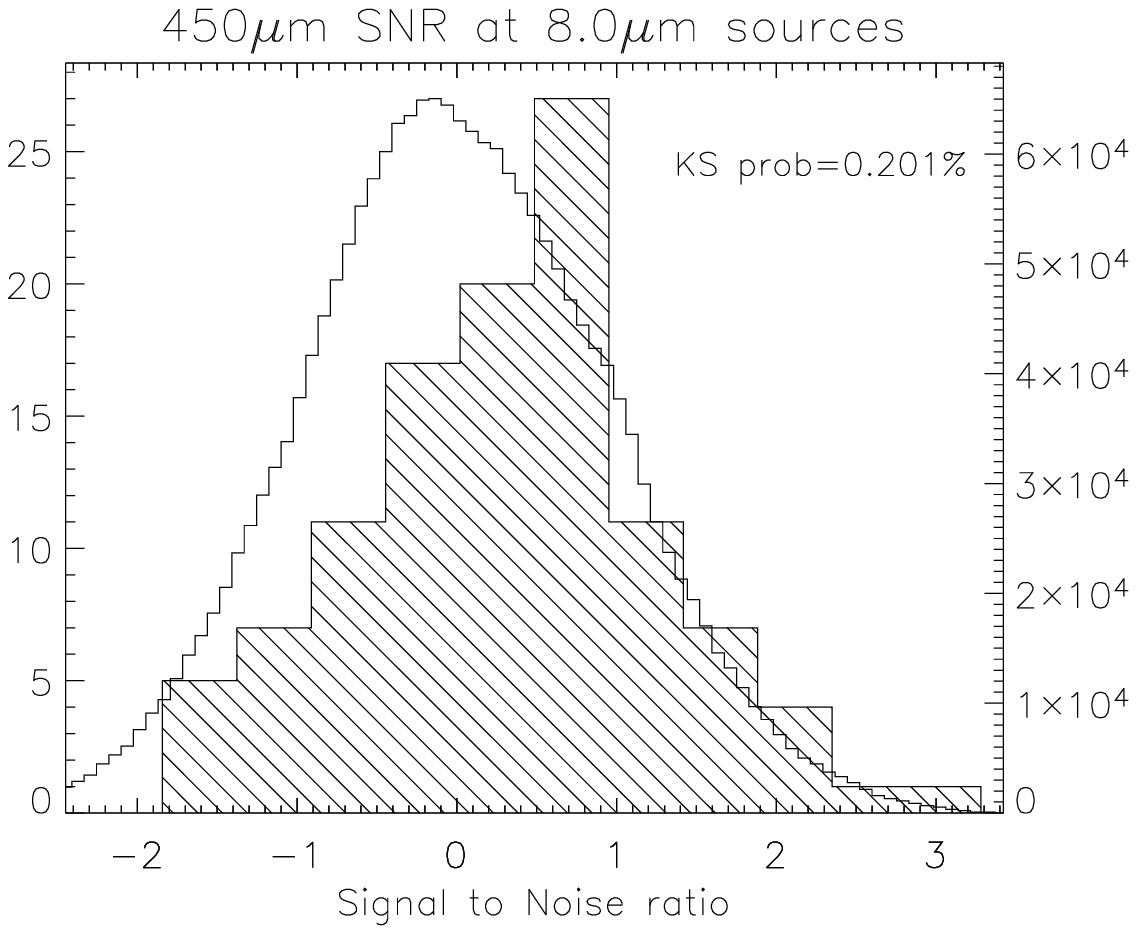}
\hspace*{-2.6cm}
\hspace*{2.5cm}
\includegraphics*[scale=0.35]{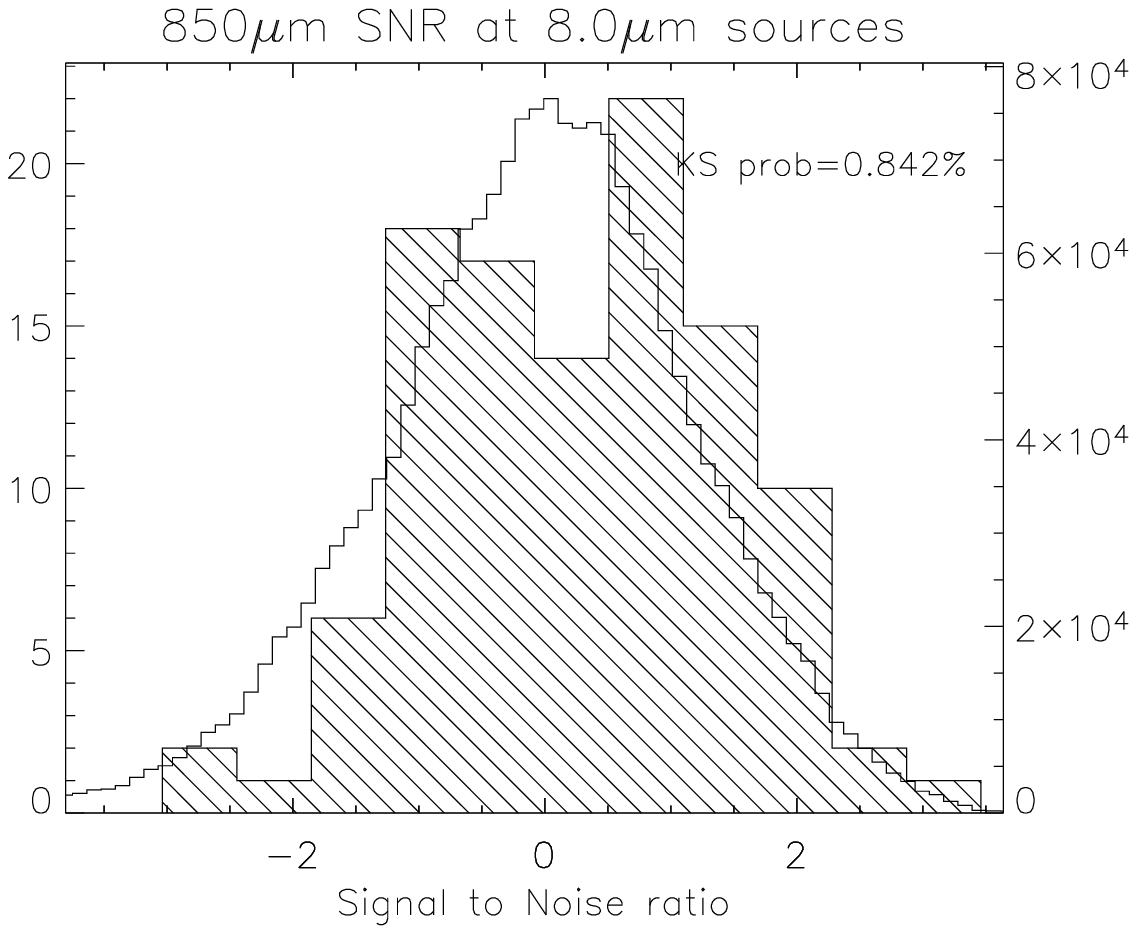}
}
\caption{\small\label{fig:histograms}
Histograms of the sub-millimeter signal-to-noise ratios (see text)
at the 
positions of the {\it Spitzer} $5.8\mu$m and $8\mu$m galaxies
(hatched histograms, left-hand ordinates), compared to the histograms
for the unmasked regions of the sub-millimeter maps as a whole
(unhatched histograms, right-hand ordinates). Note the clear positive
skew in all four hatched histograms, relative to the unhatched
histograms. The diagrams are annotated with
the probabilities that identical distributions would differ by as much
as is observed, 
according to Kolmogorov-Smirnov tests (see also table
\ref{tab:coadd_results}). 
} 
\end{figure*}

The sub-millimeter maps give the best-fit point source flux at 
each location (equation A4 of Serjeant et al. 2003a), so we 
measure the values of the sub-millimeter images at the {\it Spitzer}
galaxy positions. 
Figure \ref{fig:histograms} shows the sub-millimeter 
signal-to-noise ratios (equation A6 of Serjeant et al. 2003a)
at the positions of the {\it Spitzer}
Early Release Observations source catalogs at $5.8\mu$m and
$8\mu$m, which lie in the unmasked regions. 
These figures also show 
histograms 
for
the whole of the unmasked maps. 
Note the clear positive
skews in the sub-millimeter fluxes at the {\it Spitzer} source positions,
relative to 
the maps as a whole. We used 
Kolmogorov-Smirnov tests to determine the confidence level of these
relative positive skews, 
listed in table~\ref{tab:coadd_results}. 
There are significant detections at $5.8\mu$m and $8\mu$m, and 
marginal detections at $24\mu$m. 
No significant skew was found at $3.6\mu$m or $4.5\mu$m. 

We also calculated the mean fluxes in the sub-millimeter
maps at the positions of the {\it Spitzer} galaxies. These are almost 
all positive and are listed
in table \ref{tab:coadd_results}. This statistic is less efficient
than the Kolmogorov-Smirnov test (which does not just use the first
moment), so the uncertainties on the mean
fluxes are larger than suggested by the Kolmogorov-Smirnov
confidence levels. 
The {\it Spitzer}
galaxies are individually undetected in the sub-millimeter, 
so one must be careful in
interpreting the stacking analysis; for example, 
we do not use the {\it noise-weighted} mean
fluxes because they can give false 
positives, as noted by Serjeant et al. 2003a. 
Also, many 
fluctuations in the
sub-millimeter maps are blends (e.g. Scott et al. 2002), so is there a
risk of overestimating the flux of any given {\it Spitzer} galaxy by
also counting its neighbors? The answer is no, as Peacock et
al. (2000) showed: for a Poissonian sampling of the sub-millimeter
map, the expected total sub-millimeter flux from all neighbors equals 
the mean flux of the map, which is exactly zero for the chopped,
nodded SCUBA maps. (The chopped/nodded PSF is also zero-sum, so the
PSF-convolved images are still zero-sum.)
Blending slightly degrades the uncertainty on the
mean, but it does not affect the mean fluxes.
Another demonstration of this is to note that the total flux of any
given source is exactly zero in these chopped, nodded maps. Even
confusing sources within the same SCUBA beam will have no {\it net}
effect, provided that there are also 
similar sources uniformly distributed over the rest of the map.

The histograms of the whole map in figure \ref{fig:histograms}
effectively act as the control sample. Nevertheless, as a test of the
stacking analysis methodology, we performed the same tests on 
simulated {\it Spitzer} catalogs of the same size as the observed
catalogs, 
with a Poissonian distribution in the unmasked
regions. 
No positive skews in the sub-millimeter flux
distributions were detected relative to the maps as a whole. These
simulations also verified that the mean sub-millimeter flux
around randomly chosen positions is consistent with zero. 

\section{Discussion}

Although our {\it Spitzer} $5.8\mu$m and $8\mu$m catalogues are small
($154$ and $111$ respectively), we can use our statistical detections to 
obtain 
constraints 
on the contribution to the sub-millimeter
extragalactic background light from the {\it Spitzer} populations, by 
multiplying the mean sub-millimeter flux of a single {\it Spitzer} galaxy
(table \ref{tab:coadd_results}) with the observed surface
density of {\it Spitzer} galaxies. 
At $850\mu$m, the
integrated contribution from the $5.8\mu$m population is $0.12\pm0.05$
nW m$^{-2}$ sr$^{-1}$, and from the (overlapping) $8\mu$m population
$0.10\pm0.04$ nW m$^{-2}$ sr$^{-1}$. This is already comparable to the
total $850\mu$m extragalactic background reported by Lagache et
al. (2000) of $\sim0.5$ nW m$^{-2}$ sr$^{-1}$. At $450\mu$m the
{\it Spitzer} contributions can, within the uncertainties, account
for all the observed $\sim3$ nW m$^{-2}$ sr$^{-1}$ extragalactic 
background: $2.4\pm0.7$ nW m$^{-2}$ sr$^{-1}$ from the $5.8\mu$m-selected
sample, and $2.2\pm0.6$ nW m$^{-2}$ sr$^{-1}$ from the $8\mu$m galaxies. 
Our sub-millimeter faint source density is consistent with those in
the deepest $850\mu$m maps (Cowie et al. 2002). 

The sub-millimeter:{\it Spitzer} flux ratios in sub-millimeter point
sources (Egami et al. 2004, see above) are much higher
than average in the population (table \ref{tab:coadd_results}). 
Despite the commonality between the SCUBA and {\it Spitzer} galaxy
populations, 
sub-millimeter galaxies are not at all representative of
the {\it Spitzer} population. The sub-millimeter background
independently suggests the same result: 
if the $24\mu$m population had
sub-millimeter:{\it Spitzer} flux ratios as high as SCUBA galaxies
(e.g. median 
$S_{850\mu{\rm m}}/S_{24\mu{\rm m}}=34.4$ for the Egami et al. (2004)
sample), the $>170\mu$Jy $24\mu$m sources alone 
would over-predict the $850\mu$m background by 
$\sim60\%$. Our weak sub-millimeter detection of the $24\mu$m population
is less surprising in that context. One possible interpretation is
that the SCUBA population has fewer (or more heavily obscured) active
nuclei than the {\it Spitzer} population as a whole. 

Is our stacking analysis signal due to dusty galaxies, or to 
high-$z$
galaxies, or both? Our samples are too small to distinguish 
these possibilities, but stacking analyses of subsets show that
either option is credible. Almost all the $5.8\mu$m galaxies are detected at
$3.6\mu$m, and as with the $5.8\mu$m population as whole 
(table \ref{tab:coadd_results}) 
this subset ($153$ galaxies) is detected statistically. 
The $299$ remaining $3.6\mu$m galaxies are not detected in the stacking
analyses (e.g., only $37\%$ confidence in the $450\mu$m signal:noise
maps), so the redder galaxies are more likely to be sub-millimeter 
emitters. The lack of a stacking signal at $3.6\mu$m and $4.5\mu$m is
at least in part due to dilution of the signal from the high density
of bluer galaxies detected at these wavelengths. The red colors could
be due to heavy obscuration, but the colors also become redder quickly
with increasing redshift. 
Simpson
\& Eisenhardt (1999) show that $F_{4.5\mu\rm m}/F_{3.6\mu\rm m}>1.32$
or $F_{8\mu\rm m}/F_{5.8\mu\rm m}>1.32$ selects redshifts
$z\stackrel{>}{_\sim}1.5$. There are $47$ galaxies satisfying either
criteria, with detections in the relevant bands, and these galaxies
are detected in the stacking analysis (e.g. $99.1\%$ confidence at
$450\mu$m); the corresponding $400$ foreground 
$z\stackrel{<}{_\sim}1.5$ 
galaxies are not detected (e.g. only $71\%$ confidence at
$450\mu$m). Further work on larger samples is needed to 
discriminate between high-redshift and high-obscuration
sub-populations. 

The sub-millimeter backgrounds are sensitive to contributions
from the far-infrared luminous energy densities throughout the
redshift range of favorable K-corrections. It follows that the {\it
Spitzer} $5.8\mu$m and $8\mu$m-selected galaxies are necessarily
significant contributors to the comoving volume averaged star
formation density. 
Many of the $5.8\mu$m and $8\mu$m galaxies are challenging targets for
8-10m-class optical/near-infrared spectroscopy, but redshift surveys 
of these populations, together with 2-D and 3-D clustering, would
discriminate between competing semi-analytic descriptions of galaxy
evolution, and determine whether (or which) SCUBA/{\it Spitzer}
galaxies are the sites of the assembly of giant ellipticals. 
Finally, it is worth stressing that the conclusions in this paper 
could only have been reached with co-ordinated multi-wavelength
surveys. 

%
%
%

%
%
%
\acknowledgments{
We would like to thank the referee, Yasunori Sato, for many thoughtful
and useful comments. 
This work is based in part on observations made with the {\em Spitzer
Space Telescope}, 
(\facility{Spitzer(IRAC),Spitzer(MIPS)}, dataset
\dataset{ads/sa.spitzer 6620160,6619904})
which is operated by the Jet Propulsion Laboratory,
California Institute of Technology under NASA contract 1407. 
Support for this work was provided by NASA through Contract Number
\#960785 
and \#1256790 
issued by JPL/Caltech.
}

%
%
%

\pagebreak[4]

\clearpage

\begin{table*}
{\scriptsize
\centerline{\sc Sub-millimeter stacking analysis of {\em Spitzer}
source catalogs}
\hspace*{-1.5cm}
\begin{tabular}{llllllllllll}
\hline\hline
\noalign{\smallskip}
Wave- & 
$450\mu$m     & 
$450\mu$m & 
$450\mu$m  & 
$\langle S_{450}\rangle$ & 
$\langle S_{450}\rangle/$ & 
$850\mu$m     &
$850\mu$m     &
$850\mu$m & 
$\langle S_{850}\rangle$ & 
$\langle S_{850}\rangle/$ 
  \\
length & 
$N_{\rm src}$ & 
K-S (S:N) & 
K-S (Flux) & 
mJy & 
$\langle S_{\rm\it Spitzer}\rangle$ & 
$N_{\rm src}$ &
K-S (S:N) & 
K-S (Flux) & 
mJy & 
$\langle S_{\rm\it Spitzer}\rangle$ 
\\
\noalign{\smallskip}
\hline
\noalign{\smallskip}
$3.6\mu$m    & $452$ & $71.5\%$  & $69.3\%$  & $1.32\pm0.78$ & $38$  & $437$ & $67.3\%$  & $61.8\%$  & $0.07\pm0.10$ & $2.0$\\
$4.5\mu$m    & $436$ & $76.9\%$  & $70.1\%$  & $1.58\pm0.79$ & $64$  & $421$ & $76.2\%$  & $85.2\%$  & $0.09\pm0.10$ & $3.4$\\
$5.8\mu$m    & $154$ & $99.72\%$ & $99.80\%$ & $4.19\pm1.33$ & $95$  & $141$ & $98.66\%$ & $99.2\%$  & $0.39\pm0.17$ & $8.8$\\
$8\mu$m      & $111$ & $99.80\%$ & $99.76\%$ & $5.51\pm1.57$ & $98$  & $98$  & $99.16\%$ & $99.65\%$ & $0.48\pm0.20$ & $9.4$\\
$24\mu$m$^a$ &  $83$ & $82.5\%$  & $84.6\%$  & $2.24\pm2.08$ & $9.3$ & $78$  & $98.42\%$ & $97.65\%$ & $0.30\pm0.24$ & $1.25$\\
$24\mu$m$^b$ &  $42$ & $88.3\%$  & $91.51\%$ & $4.35\pm2.92$ & $12.9$& $38$  & $97.87\%$ & $97.32\%$ & $0.55\pm0.35$ & $1.68$\\
%
%
%
\noalign{\smallskip}
\hline
\end{tabular}
\caption{
\label{tab:coadd_results}
\scriptsize
Stacking analysis of the {\it Spitzer} galaxy populations in the
sub-millimeter maps. 
The $24\mu$m $>120\mu$Jy and $>170\mu$Jy
catalogs are listed separately ($^a$ and $^b$ respectively). 
The $N_{\rm src}$ values are the numbers of {\it
Spitzer} sources used in each stacking analysis. 
The Kolmogorov-Smirnov test results quoted are
confidence levels, i.e. a high number indicates a net statistical
detection of {\it Spitzer} galaxies. 
Results are listed for 
the signal-to-noise histograms
(plotted in figure \ref{fig:histograms}) 
as well as 
the flux distributions. 
At $5.8\mu$m and $8\mu$m the
galaxies are detected, marginally so at $24\mu$m, 
but not at $3.6\mu$m or $4.5\mu$m. 
Also given are mean sub-millimeter fluxes, and
mean flux 
ratios, at the positions of the {\it Spitzer} sources. 
Errors quoted are $\sigma/\sqrt{N}$, where $N$ is
the number of measurements and $\sigma^2$ the variance of the
measurements. 
Known sub-millimeter sources have been excluded from this
analysis, as discussed in the text.
}
}
\end{table*}

\end{document}